\documentclass[useAMS,natbib]{mn2e}
\usepackage{amssymb,amsmath}
\usepackage{graphicx}

\title[Accretion Rate of AGN]
{The Correlation Between Spectral Index And Accretion Rate For AGN}

\author[Zhang et al.]
      {Xue-Guang Zhang$^{1,3}$\thanks{xueguang@mpa-garching.mpg.de},
       Deborah Dultzin$^1$,
       Ting-Gui Wang$^2$ \\
      $^1$Instituto de Astronom\'ia, Universidad Nacional Aut\'onoma de
                M\'exico, Apdo Postal 70-264, M\'exico D. F. 04510, Mexico \\
      $^2$Center for Astrophysics, Department of astronomy and Applied
                Physics, University of Science and Technology of China, \\
                Hefei, Anhui, P.R.China\\
      $^3$Max-Plank Institute f\"ur Astrophysik, Karl-Schwarzschild-str 1,
          85748 Garching, Germany}

\date{}

\def\LaTeX{L\kern-.36em\raise.3ex\hbox{a}\kern-.15em
   T\kern-.1667em\lower.7ex\hbox{E}\kern-.125emX}
\begin{document}
\label{firstpage}

\maketitle
\begin{abstract}
In this paper, we present a correlation  between the  spectral 
index distribution (SED) and the dimensionless accretion rate 
defined as $\dot{m}={L_{bol}/L_{Edd}}$ for AGN.
This quantity  is used as a substitute of the physical accretion rate.
We select 193 AGN with both broad H$\alpha$ and broad H$\beta$,
and with absorption lines near MgI$\lambda5175\AA$ from SDSS DR4. We
determine the spectral index and dimensionless accretion
rate after correcting for both host galaxy contribution and internal
reddening effects. A correlation is found between the optical
spectral index and the dimensionless accretion rate for AGN, including
low luminosity AGN ($L_{H\alpha}<10^{41}{\rm erg\cdot s^{-1}}$
sometimes called "dwarf AGN" (Ho et al. 1997)).
The existence of this correlation provides an independent
method to estimate the central BH  masses for all types of AGN.
We also find that there is a different correlation between the spectral
index and the BH masses for normal AGN and low luminosity AGN, which
is perhaps due to the different accretion modes in these two types of
nuclei. This in turn may lead to the different correlations between BH
masses and optical continuum luminosity reported previously
(Zhang et al. 2007a), which invalidates the application of the empirical
relationship found by Kaspi et al. (2000, 2005)
to low luminosity AGN in order to determine their BLR sizes.
\end{abstract}


\section{Introduction}

    Because of the difficulty to calculate the
physical accretion rate to the BH in an AGN, a dimensionless accretion 
rate can be defined and estimated based on the bolometric luminosity and 
BH mass, $\dot{m} = \frac{L_{bol}}{L_{Edd}}$. This parameter
can be used as the substitute for the actual accretion rate.
Moreover, the dimensionless accretion rate is an important parameter
in the scheme of the so called unified model for AGN (Antonucci 1993,
Quintilio \& Viegas 1997, Urry \& Padovani 1995), and also the 4D Eigenvector
1 Scheme (e.g., Dultzin-Hacyan et al. (2007).
The difference in  accretion rate leads to the principal difference
between high luminosity QSOs and low luminosity Seyfert galaxies. It seems
to be also one of the main physical parameters underlying
the 4D Eigenvector 1 Scheme as explained recently in
Dultzin-Hacyan et al. (2007). In order to obtain the dimensionless accretion
rate $\dot{m}$, two other parameters must be calculated first:
BH mass $M_{BH}$ and the bolometric luminosity $L_{bol}$.

   The common method to estimate the bolometric luminosity $L_{bol}$
of AGN is based on the continuum luminosity from the nuclei:
$L_{bol}\sim 9\times L_{5100\AA}$ given by Kaspi et al.
(2000) and confirmed by Shang et al. (2005) for QSOs. Recently,
the relation was used by Bonning et al. (2007) to study the
correlation between the accretion disk temperatures and the
continuum colors in QSOs. However, it should be stressed  that the relation
does not hold for ALL kinds of AGN, in particular, for low luminosity
AGN (Ho et al. 1997a, 1997b, Ho 1999), because of the different Spectral
Energy Distribution (SED) (the lack of the Big Blue Bump). 

   Several methods are used to estimate BH masses of AGN.
The most reliable method is based on the stellar velocity dispersion
of the bulge of the host galaxy first presented by
Ferrarese \& Merritt (2000) and Gebhardt et al. (2000), then confirmed by
Tremaine et al. (2002) and Merritt \& Ferrarese (2001)
etc.
\begin{equation}
M_{BH} = 10^{8.13\pm0.06}(\frac{\sigma}{200{\rm km\cdot s^{-1}}})^{4.02\pm0.32} {\rm M_{\odot}}
\end{equation}
which indicates a strong correlation between BH masses and bulge masses
(H\"{a}ing \& Rix 2004, Marconi \& Hunt 2003, McLure \& Dunlop 2002,
Laor 2001, Kormendy 2001, Wandel 1999) etc.
However we should note that the relation of $M_{BH} - \sigma$ is
obtained through the results of nearby inactive galaxies. Whether the
relation can be applied to far away active galaxies is an interesting
question. So far, there are a few dynamical mass estimates of central black
holes of broad line AGN, and the BH masses are consistent with the BH masses
estimated from the relation of $M_{BH} - \sigma$, although the uncertainties
are still large. In addition, we should say that the objects in our sample
described in the following section are not high luminosity and high redshift
QSOs, thus the correlation between central black hole and bulge of host
galaxy can be reasonably considered to hold.

   The other methods are based on the assumption of virialization of the
Broad Line Emitting Regions (BLRs),
or at least part of them (Peterson et al. 2004, Onken et al. 2004,
Sulentic et al. 2006, Dultzin-Hacyan et al. 2007). In order to calculate
the parameter of dimensionless accretion rate, the  BH mass is
necessary. However, for high luminosity and high redshift AGN, it is
difficult to measure the stellar velocity dispersions. The assumption
of virialization is applied for QSOs. In order to estimate BH masses
of QSOs based on the assumption of virialization, the most convenient
way is to use the equation:
\begin{equation}
\begin{split}
M_{BH} &= f\times\frac{R_{BLRs}\times\sigma_{b}^2}{G} \\
       &= 2.15\times10^8(\frac{\sigma_b}{3000{\rm km\cdot s^{-1}}})^2(\frac{L_{5100\AA}}{10^{44}{\rm erg\cdot s^{-1}}})^{0.69} {\rm M_{\odot}}
\end{split}
\end{equation}
There are, however, some caveats with this method. First, the question of whether the
relation $R_{BLRs}\sim L_{5100\AA}^{0.69}$ found by Kaspi et al.
(2000, 2005) can be applied for all AGN, in particular high redshift
ones. In an attempt to answer this question,
we have found that the relation is not valid for some special
kinds of AGN, such as the low luminosity AGN (Zhang, Dultzin-hacyan \&
Wang 2007a, Wang \& Zhang 2003) and the AGN with double-peaked low ionization
emission lines (Zhang, Dultzin-Hacyan \& Wang 2007b).  Second, the
estimation of the BH masses of high redshift AGN by means of Equation (2)
will lead to BH masses larger than $10^{10}{\rm M{\odot}}$ (meaning
$\sigma>600{\rm km\cdot s^{-1}}$), which leads to  unreasonable
masses of the bulge larger than $10^{13}{\rm M_{\odot}}$ (Netzer 2003,
Sulentic et al. 2006). For this reason, finding another parameter which can
be observationally determined, related to the dimensionless accretion rate
is an important task and is the main objective of this paper.
The accretion rate is determined by two properties: the continuum
luminosity $L_{5100\AA}$ and the BH mass $M_{BH}$. The continuum
luminosity can be calculated from the observed spectra as discussed in
the the next section. Thus, in order to obtain a reliable result, we select
Equation (1) to estimate the central BH masses of AGN rather than Equation (2).

   The accretion disk model has been widely accepted as the standard model
for AGN. In the NLTE (Non Local Thermodynamic Equilibrium) accretion disk
mode, the generated SED (Spectral Energy Distribution) is based on three
main parameters: BH masses $M_{BH}$, accretion rate $\dot{M}$ and the
viscosity parameter $\alpha$. An expected result is that there should be a
correlation between the spectral index and the accretion
rate $\dot{m}$. In this paper, we answer the question whether the
observed spectral index can be used to trace the  dimensionless
accretion rate. In section II, we present the data sample. Section III
gives the results. Finally  the discussion and conclusions are given in
Section IV. In this paper, the cosmological parameters
$H_{0}=70{\rm km\cdot s}^{-1}{\rm Mpc}^{-1}$, $\Omega_{\Lambda}=0.7$ and
$\Omega_{m}=0.3$ have been adopted.

\section{Data Sample}

We select objects from SDSS DR4 (Adelman-McCarthy et al. 2006) to make up
our sample according to the following two criteria: First, and most
important is that the objects' spectra present absorption features, here we
focused on the absorption line MgI$\lambda5175\AA$, in order to measure the
stellar velocity dispersion of the bulge. Second, in order to obtain the
intrinsic continuum luminosity from the nuclei after the correction of
internal reddening effects using the Balmer decrement,  Balmer
emission lines, both H$\alpha$ and H$\beta$ must be also present.
In order to perform accurate measurements of the lines mentioned
above, several procedures have to be followed.

   In order to obtain the continuum luminosity from the nuclei we must
subtract first the contribution of stellar light. An efficient method to
subtract the stellar light is the PCA (Principle Component analysis) method
described by Li et al. (2005) and Hao et al. (2005), using the eigenspectra
from pure absorption galaxies from SDSS or the eigenspectra from stars
in STELIB (Le Borgne et al. 2003), because the method of Principle Component
Analysis (PCA) provides a better way to constrict more favorable
information from a series of spectra of stars or galaxies into several
eigenspectra. Here, we used the method from
Hao et al. (2005). The eigenspectra are calculated by KL (Karhunen-Loeve)
transformation for about 1500 pure absorption galaxies selected from SDSS
DR4. Then, the first eight eigenspectra and the spectra of an A star (which
is used to account for star formation)
selected from STELIB (Le Borgne et al. 2003) are used to fit the
stellar properties of the observed spectra. After this,  rather
than a power law, a three-order polynomial function is used to fit the
featureless continuum, because the study of composite spectra of AGN shows
that the continuum should be best fitted by two power laws with a
break of $\sim5000\AA$ (Francis et al. 1991, Zheng et al. 1997,
Vanden Berk et al. 2001). After the last step, the featureless continuum and
the stellar components are obtained based on the Levenberg-Marquardt
least-squares minimization method.

   After the subtraction of stellar components and the continuum
emission, the line parameters of emission lines can be measured by
Levenberg-Marquardt least-squares minimization: one gaussian function for
each forbidden emission line, two gaussian functions (one broad and one
narrow) for each permitted emission line. For [OIII]$\lambda4959,5007\AA$,
we use an extra gaussian function  for the extended wings as shown
in Greene \& Ho (2005a). Then we select the objects with reliable broad
H$\alpha$ and broad H$\beta$ according to the following criteria:
$\sigma(B)\ge3\times\sigma(B)_{err}$, $flux(B)\ge3\times flux(B)_{err}$
and $\sigma(B)\ge600{\rm km\cdot s^{-1}}$, where 'B' represents the values
for the broad Balmer components, 'err' means the measured error of the
value, $\sigma$ is the second moment of broad Balmer emission lines.

     Then it is necessary  to measure the stellar velocity
dispersions of the objects selected. However, the accurate measurement
of stellar velocity dispersion is an open question, because of the known
problems with the template mismatch. A commonly used method is to select
spectra of several kinds of stars (commonly, G and K) as templates, and
then broaden the templates by the same velocity to fit stellar features,
leaving the contributions from different kinds of stars as free parameters
(Rix \& White 1992). However, more information about stars included
by the templates should lead to more accurate measurement of stellar
velocity dispersion. According to the above mentioned method to subtract
stellar components, we created a new template rather than several spectra
of G or K stars as templates. Thus, we apply the PCA method for all 255
spectra of different kinds of stars in STELIB. Selecting the first several 
eigenspectra and a three-order polynomial function for the background as 
templates, the value of stellar
velocity dispersion can be measured by the minimum $\chi^2$ method applied to the absorption features around MgI$\lambda5175\AA$ within the wavelength range from 5100$\AA$ to
5300$\AA$ (Zhang, Dultzin-Hacyan \& Wang 2007c). Finally, we select
the objects for which the measured values of stellar velocity dispersions are at least three
times larger than the measured errors.

   Finally, we select 193 AGN with redshift from 0.015 to 0.25 and
with observed featureless continuum luminosity within the range from
$10^{41.23} {\rm erg\cdot s^{-1}}$ to $10^{43.79} {\rm erg\cdot s^{-1}}$
from about 400000 objects classified as
galaxies in SDSS DR4. The objects have reliable stellar velocity
dispersions and reliable broad Balmer emission lines.

\section{Results from the database of SDSS}

   Before proceeding further, a simple discussion about the origin
of the featureless continuum emission is given. Basically, the
possibility of nebular emission can be rejected. According to
the luminosity of Recombination lines, the nebular continuum emission
at $5100\AA$ can be simply estimated by
$L_{5100\AA, Nebulae}\sim0.1\times L_{H\beta}$, if the electron
temperature $T = 10^4{\rm K}$ is accepted. Thus the effects of nebular
emission can be neglected. Furthermore, we check the correlation
between continuum luminosity and the luminosity of Balmer emission
lines found by Greene \& Ho (2005b). The result is shown in Figure 1.
We should note that the continuum luminosity and luminosity of 
H$\alpha$ are the values before the internal reddening correction as 
shown in Greene \& Ho (2005b).
The coincident correlation between $L_{5100\AA}$ and $L_{H\alpha}$ (
including the narrow component) indicates that our method to subtract
the stellar components is reliable to some extent. Based on the correlation of
$L_{H\alpha} - L_{5100\AA}^{1.157}$ for AGN with high continuum
luminosity, we can estimate the effects of star
formation on the continuum luminosity. For AGN with low luminosity and
low redshift, there are two components in narrow H$\alpha$, one from the AGN 
$L_{H\alpha,AGN}$ and the other one from star formation $L_{H\alpha,SF}$
(Kauffmann et al. 2003). Moreover, we assumed the continuum luminosity
also includes two components, one from the AGN $L_{5100\AA,AGN}$ and the
other one from star formation $L_{5100\AA,SF}$. Furthermore, there is 
a strong correlation between line and continuum luminmosities shown in Figure 1, $L_{H\alpha,SF}+L_{H\alpha,AGN}\propto(L_{5100\AA,SF}+L_{5100\AA,AGN})^{1.157}$. We can simply estimate
the effects of star formation on continuum luminosity, if we accept
that $L_{H\alpha,SF} = s\times L_{H\alpha,AGN}$ and
$L_{H\alpha,AGN}\propto L_{5100\AA,AGN}^{1.157}$ (because the relation is
better applied for high luminosity AGN with less effects of star formation):
\begin{equation}
\begin{split}
&L_{H\alpha,SF}+L_{H\alpha,AGN}\sim(L_{5100\AA,SF}+L_{5100\AA,AGN})^{1.157} \\
&L_{H\alpha,AGN}\sim L_{5100\AA,AGN}^{1.157} \\
&1 + s \sim (1 + \frac{L_{5100\AA,SF}}{L_{5100\AA,AGN}})^{1.157}
\end{split}
\end{equation}
If we accepted that star-forming regions contribute 65\% of the narrow
H$\alpha$ flux as described in Kauffmann et al. (2003), the parameter of
$s$ can be determined as:
\begin{equation}
\begin{split}
s &= \frac{L_{H\alpha,SF}}{L_{H\alpha,AGN}} \\
  &= \frac{0.65\times L_{H\alpha,N}}{L_{H\alpha,B}+0.35\times H_{H\alpha,N}}
\end{split}
\end{equation}
where 'N' and 'B' represent the narrow component and broad component of
H$\alpha$. The mean value of $L_{H\alpha,B}/L_{H\alpha,N}$ is about 3.96
for the objects in our sample. Then we can determine that
$L_{5100\AA,SF}/L_{5100\AA,AGN}\sim0.12$, i.e., the star-forming regions
contribute about 10\% of the observed continuum luminosity. Thus
in the following section, we can ignore the effects of star fromation.

   In order to obtain a reliable intrinsic continuum shape, the
internal reddening effects must be corrected. The common way
to correct them  is through the Balmer
decrement. Here, we assume the intrinsic Balmer decrement as 3.1
for  H$\alpha$ and H$\beta$ expected by Case B recombination (albeit it is debatable wether it
can be applied to broad lines) with
some contributions from collisional excitation.
Then, the value of E(B-V) can be determined from the Balmer
decrement:
\begin{equation}
{\rm E(B-V)} = -0.97615448+1.9866313\log(\frac{{\rm H}\alpha}{{\rm H}\beta})
\end{equation}
where $\frac{{\rm H}\alpha}{{\rm H}\beta}$ is the observed flux ratio.
This equation calculated from the R-dependent Galactic extinction
curve presented by Fitzpatrick (1999) can be used to calculate the
value of E(B-V) simply through the Balmer decrement. In the
following of the paper, the continuum luminosity and luminosity of H$\alpha$
are the ones after the correction of BLRs extinction.

    After the correction of internal reddening effects, the spectral
index can be determined. Here we select three spectral indices:
$\frac{F_{4400\AA}}{F_{5100\AA}}$, $\frac{F_{5100\AA}}{F_{6800\AA}}$ and
$\frac{F_{4400\AA}}{F_{6800\AA}}$. The BH masses are calculated using
Equation (1). The internal continuum luminosity after the correction of
the internal reddening  can also be calculated. Thus it is
easy to check the correlation between dimensionless accretion rate
and spectral index. Here we should notice that the diminsionless accretion
rate for low luminosity AGN as discussed in the introduction and in the next
section is also calculated by
$\dot{m}=\frac{L_{bol}}{L_{Edd}}\sim\frac{9\times L_{5100\AA}}{L_{Edd}}$,
although, the bolometric luminosity of low luminosity AGN cannot be correctly
calculated by $L_{bol}\sim9\times L_{5100\AA}$ as shown in Ho (1999).
However, to some extent, we can accept that, if there is also a simple relation 
$L_{bol}\sim k\times L_{5100\AA}$ for low luminosity AGN, the calculated
$\frac{9\times L_{5100\AA}}{L_{Edd}}$ can be used as a substitute
of accretion rate, and it is convenient to compare the properties of
low luminosity and normal AGN.
The correlations are shown in Figure 2. The Spearman
Rank Correlation Coefficient is 0.68 with $P_{null}\sim1.51\times10^{-27}$,
0.65 with $P_{null}\sim1.42\times10^{-24}$ and
0.67 with $P_{null}\sim1.94\times10^{-26}$ for
$\frac{F_{5100\AA}}{F_{6800\AA}}$, $\frac{F_{4400\AA}}{F_{5100\AA}}$ and
$\frac{F_{4400\AA}}{F_{6800\AA}}$ respectively. In order to check
the effects of internal reddening, we also show the correlation
between the dimensionless accretion rate and the spectral index
$\frac{F_{5100\AA}}{F_{6800\AA}}$ in the right-bottom panel of Figure 2,
without internal reddening correction. The Spearman Rank
Correlation Coefficient for the correlation without reddening correction
is about 0.56 with $P_{null}\sim2.01\times10^{-17}$.
The unweighted best fitted results for the correlations between spectral
indices and dimensionless accretion rates are shown as solid lines that correspond to:
\begin{equation}
\begin{split}
&\log(\frac{F_{5100\AA}}{F_{6800\AA}}) = 0.56 + 0.18\times\log(\frac{9\times L_{5100\AA}}{L_{Edd}}) \\
&\log(\frac{F_{4400\AA}}{F_{5100\AA}}) = 0.35 + 0.14\times\log(\frac{9\times L_{5100\AA}}{L_{Edd}}) \\
&\log(\frac{F_{4400\AA}}{F_{6800\AA}}) = 0.92 + 0.33\times\log(\frac{9\times L_{5100\AA}}{L_{Edd}}) \\
&\log(\frac{F_{5100\AA}}{F_{6800\AA}}) (uncorr)= 0.37 + 0.14\times\log(\frac{9\times L_{5100\AA}}{L_{Edd}})
\end{split}
\end{equation}
The last expression is for the correlation between spectral index
$\frac{F_{5100\AA}}{F_{6800\AA}}$ and dimensionless accretion rate without
internal reddening correction. Furthermore, we are interested in the
absolute scatter in the parameter of spectra index which can be calculated
by:
\begin{equation}
\Delta_Y = \sqrt{\frac{\sum_{i=1}^{N}(Y_i-Y_{i,fit})^2}{N}}
\end{equation}
where $Y_i, Y_{i,fit}$ are the measured value of spectra index and
the fitted value by the equations listed in Equation (6).
Finally, we can obtain the scatters as follows,
$\Delta_{\log(\frac{F5100\AA}{F6800\AA})}\sim0.128$,
$\Delta_{\log(\frac{F4400\AA}{F5100\AA})}\sim0.124$,
$\Delta_{\log(\frac{F4400\AA}{F6800\AA})}\sim0.227$ and
$\Delta_{\log(\frac{F5100\AA}{F6800\AA})(uncorr)}\sim0.123$

We also show the correlation between the line width of broad
H$\alpha$ and the line width of broad H$\beta$ in Figure 3.  The Spearman
Rank Correlation Coefficient is about 0.82 with $P_{null}\sim0$.
The correlation between the line widths of broad Balmer emission lines
is:
\begin{equation}
\sigma_{H\beta_B} = (1096.56\pm124.74)\times(\frac{\sigma_{H\alpha_B}}{10^3 {\rm km\cdot s^{-1}}})^{1.01\pm0.02} {\rm km\cdot s^{-1}}
\end{equation}
where $\sigma$ is the measured value using a gaussian function to measure broad
emission lines. This correlation is similar to the one for QSOs found
by Greene \& Ho (2005b), $FWHM_{H\beta}\propto FWHM_{H\alpha}^{1.03\pm0.03}$.
This indicates that the measurement of line parameters of broad Balmer emission
lines is reliable. In addition, it is convenient for us to compare
the two kinds of BH masses estimated from Equation (1) and Equation (2).
Before proceeding further, we should notice that Equation (2) cannot
be applied to the low luminosity AGN discussed in the next section, because
of the unreasonable correlation
between the size of BLRs and the continuum luminosity found by Kaspi et al.
(2000, 2005) (Wang \& Zhang 2003, Zhang, Dultzin-Hacyan \& Wang 2007a).
Thus, here, we select the normal AGN in our sample to estimate the virial BH masses with Equation (2). Finally there are 155 objects shown in Figure 4 to
compare the two kinds of BH masses estimated by Equation (1) and Equation (2).
The Spearman Rank correlation coefficient is about 0.36 with $P_{null}\sim4.8\times10^{-6}$,
after the internal reddening corrections. The correlation also
indicates that the measured stellar velocity dispersions, the measured
internal continuum luminosities and the line widths are reliable.

\section{Discussion and Conclusions}

   There are 38 low luminosity AGN with
$L_{H\alpha}<10^{41}{\rm erg\cdot s^{-1}}$ (Ho et al. 1997a, 1997b
and Ho 1999), which are shown in solid circles in Figure 2. From the
figure, we can see that there is no difference in the correlation between
spectral index and $\frac{9\times L_{5100\AA}}{L_{Edd}}$ for normal AGN
and low luminosity AGN. If the bolometric luminosity of low luminosity
AGN was different from $9\times L_{5100\AA}$, all the low luminosity AGN
would deviate from the correlation for normal AGN, due (probably) to 
different accretion modes. Even if there is a different accretion mode for
low luminosity AGN (as suggested by the lack of the big blue bump in the spectra of low
luminosity AGN as shown in Ho (1999)), the bolometric luminosity
of low luminosity AGN can also be calculated using
$L_{bol}\sim k\times L_{5100\AA}$. Otherwise, we could not find the same
correlation between spectral index and $\frac{9\times L_{5100\AA}}{L_{Edd}}$
for low luminosity AGN and normal AGN.

  According to the accretion disk model, the output SED is the result
of the convolution of other parameters as well, such as the central
BH mass, the viscosity in the disk, and the inclination angle. However
there is no correlation between the spectral index and  central BH
masses, which is shown in Figure 5. The Spearman Rank
Correlation Coefficient is less than 0.1 with $P_{null}>60\%$ for
all objects in our sample. An interesting result is that
there is actually a negative trend (anticorrelation)  between BH masses
and the spectral indexes for the 38 low luminosity AGN.
The coefficient is about -0.54 with $P_{null}\sim4.97\times10^{-4}$,
-0.52 with $P_{null}\sim8.62\times10^{-4}$ and -0.55 with
$P_{null}\sim3.98\times10^{-4}$ for $\frac{F_{5100\AA}}{F_{6800\AA}}$,
$\frac{F_{4400\AA}}{F_{5100\AA}}$ and $\frac{F_{4400\AA}}{F_{6800\AA}}$
respectively.

   Because of the positive correlation between the spectral
index and the accretion rate, a negative correlation between the
spectral index and  central BH masses could be expected for all AGN.
However our results indicate that this expectation is only valid
for low luminosity AGN. The reason is probably related to the correlation
between BH masses and  continuum luminosity. For
Normal AGN, there is strong correlation between the  BH masses
and the continuum luminosity (Peterson et al. 2004). However for low
luminosity AGN, this correlation is much weaker (Zhang, Dultzin-Hacyan \& Wang
2007a). The correlation between the central BH masses and
the internal continuum luminosity is shown in Figure 6.
The coefficient is about 0.47 with $P_{null}\sim9.05\times10^{-10}$,
however, the coefficient is only 0.12 with $P_{null}\sim49\%$ for
the 38 low luminosity AGN.  The same result for low luminosity AGN can
be found in Panessa et al. (2006). In their paper, they selected all the
low luminosity Seyfert galaxies from Ho, Filippenko \& Sargent (1997a, 1997b),
and found that there is NO correlation between the X-ray or optical emission
line luminosities (especially [OIII]$\lambda5007\AA$ line) and BH masses.
The result also confirms that there is a different
accretion mode for normal and low luminosity AGN.

  To estimate the effects of the inclination angle of the accretion disk
is difficult. However under the assumption that the narrow line emission region
is isotropic, we can check the correlation between the continuum luminosity
and the luminosity of narrow emission lines. If the objects have very
different inclination angles of the accretion disk, a loose correlation
between the continuum luminosity and the luminosity of narrow emission
line should be expected. Here we show the correlation between $L_{5100\AA}$
and the luminosity of narrow H$\alpha$ in Figure 7. Although it is more
common to use the [OIII]$\lambda5007\AA$ as an isotropic estimator of AGN
luminosity, we prefer to use the narrow component of H$\alpha$ because of the
following reason.
[OIII] emission line frequently cannot be fitted by one single
gaussian function, because it has extended wings (Greene \& Ho 2005a).
The two components are not emitted from the same region. The
extended component is probably emitted from the far-side of the BLRs.
Thus when we fit the
[OIII] line, two gaussian functions are applied as described in Section II.
We thus select the narrow component of H$\alpha$ rather than
[OIII] to test the effects of inclination angle. A strong correlation
can be confirmed. The spearman Rank Correlation Coefficient is about
0.89 with $P_{null}\sim0$ for normal AGN, and about 0.67 with
$P_{null}\sim4.04\times10^{-26}$ for the 38 low luminosity AGN.
The best fit to the correlation (after considering the error in the determination
of the luminosity of narrow H$\alpha$) is given by:
\begin{equation}
\log(L_{H\alpha_N}) = (1.373\pm0.032) + (0.915\pm0.003)\times\log{L_{5100\AA}} {\rm erg\cdot s^{-1}}
\end{equation}
This result indicates that the effects of the inclination angle can be
neglected for the correlation between the spectral index and the
dimensionless accretion rate.

    The correlation between spectral index and dimensionless
accretion rate found in this research provides another independent
method to estimate the central BH masses of AGN. The spectral index and
continuum luminosity can be directly determined from the observed
spectrum in the optical band, and subsequently the Eddington Luminosity,
i.e. BH masses, can be determined by means of the correlation we found. This method
has the advantage of being  independent of the different correlations between
the size of the BLRs and the continuum luminosity in Equation (2). Also, this method
can be applied when it is not possible to measure
the stellar velocity dispersion of the bulge. In future work, we will
estimate the BH masses of QSOs with higher redshift using this method
to solve the problem of why  virial BH masses of QSOs estimated
by Equation (2) lead to BH masses larger than $10^{10} {M_{\odot}}$, while observational results
(fortunately) seem to contradict this result (Dultzin-Hacyan et al. 2007).

  Finally, a simple summary is as follows. We first select 193 AGN with both
broad H$\alpha$ and broad H$\beta$, and with apparent absorption
MgI$\lambda5175\AA$ from SDSS DR4. Then after the determination of
the spectral index (after the correction of internal reddening effects
through the Balmer decrements for broad Balmer emission lines, and after
the subtraction of stellar component) and
dimensionless accretion rate ($\frac{9\times L_{5100\AA}}{L_{Edd}}$),
we find a strong correlation between these parameters for AGN, which
provides another independent and method to estimate the central BH masses
of AGN.

\section*{Acknowledgements}
ZXG gratefully acknowledges the postdoctoral scholarships offered by la
Universidad Nacional Autonoma de Mexico (UNAM). D. D. acknowledges
support from grant IN100507 from PAPIIT, DGAPA, UNAM. This paper has
made use of the data from the SDSS projects.
Funding for the creation and the distribution of the SDSS Archive
has been provided by the Alfred P. Sloan Foundation, the
Participating Institutions, the National Aeronautics and
Space Administration, the National Science Foundation,
the U.S. Department of Energy, the Japanese Monbukagakusho, and the
Max Planck Society. The SDSS is managed by the Astrophysical Research
Consortium (ARC) for the Participating Institutions. The Participating
Institutions are The University of Chicago, Fermilab, the
Institute for Advanced Study, the Japan Participation Group,
The Johns Hopkins University, Los Alamos National Laboratory,
the Max-Planck-Institute for Astronomy (MPIA),
the Max-Planck-Institute for Astrophysics (MPA), New
Mexico State University, Princeton University, the United
States Naval Observatory, and the University of Washington.

\onecolumn
\begin{figure}
\centering\includegraphics[width=140mm,height = 100mm]{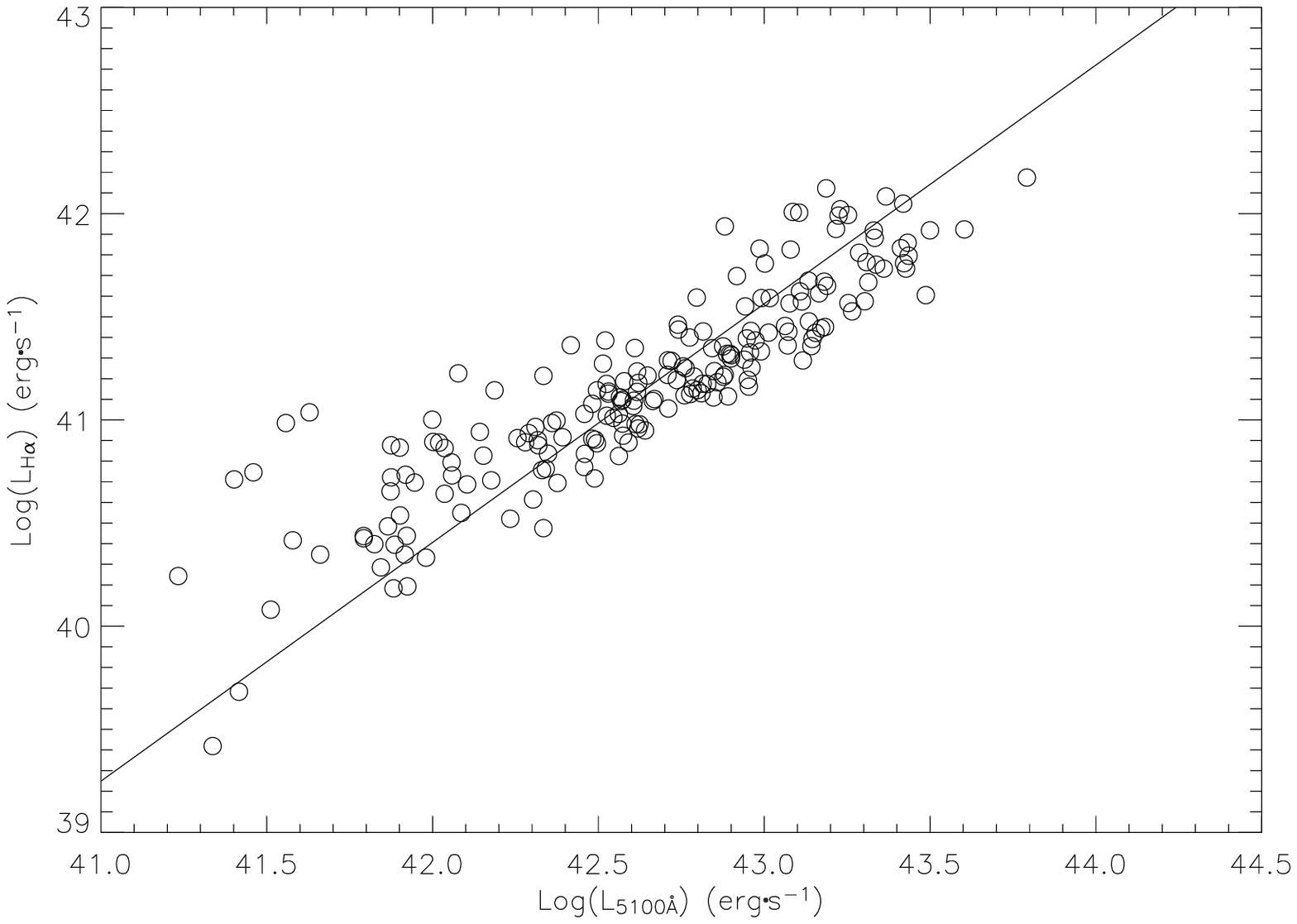}
\caption{The correlation between the continuum luminosity and the luminosity
of H$\alpha$ before the internal reddening correction.
The solid line represents the correlation found by Greene \& Ho (2005b).}
\end{figure}

\begin{figure}
\centering\includegraphics[width=140mm,height = 100mm]{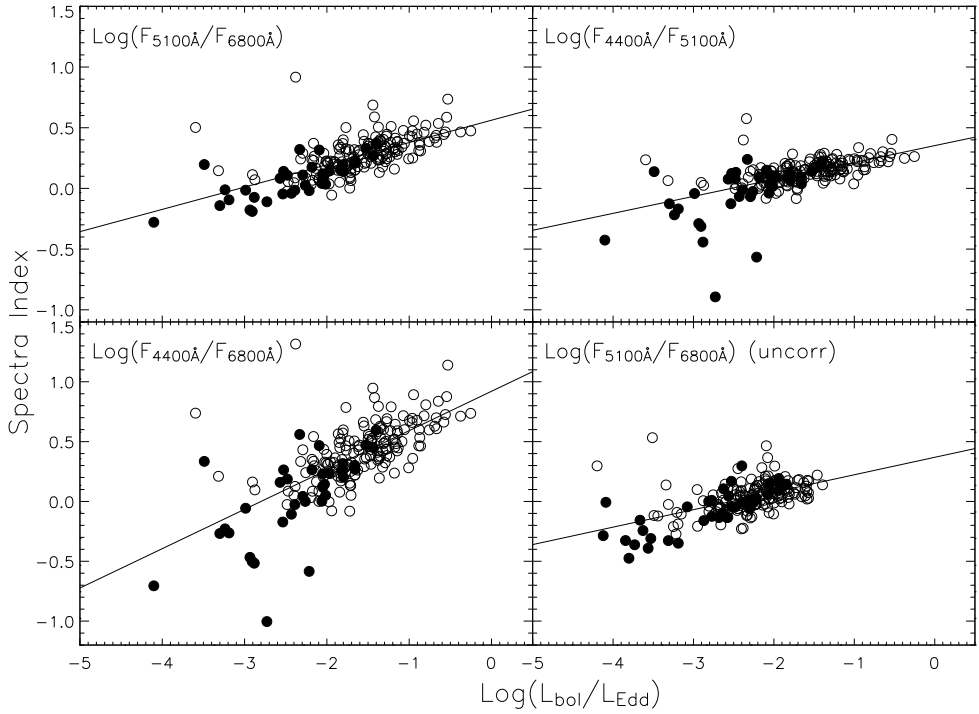}
\caption{The correlation between the spectral index and the dimensionless
accretion rate. The solid line represents the unweighted best fitted result.
Solid circles represent low luminosity AGN with luminosity of H$\alpha$ less than $10^{41} {\rm erg\cdot s^{-1}}$ as identified in
Ho et al. (1997a, 1997b), open circles are  normal AGN.}
\end{figure}

\begin{figure}
\centering\includegraphics[width=140mm,height = 100mm]{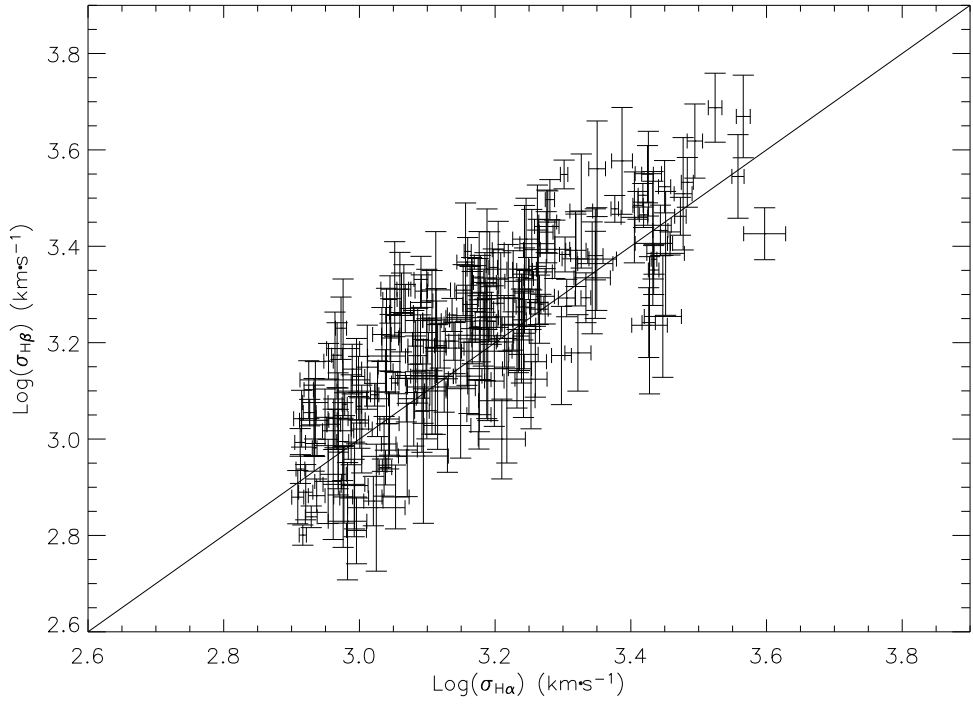}
\caption{The correlation between the line widths of broad H$\alpha$ and
 broad H$\beta$. The solid line represents the relation:
$\sigma_{H\alpha}(B) = \sigma_{H\beta}(B)$.
}
\end{figure}

\begin{figure}
\centering\includegraphics[width=140mm,height = 100mm]{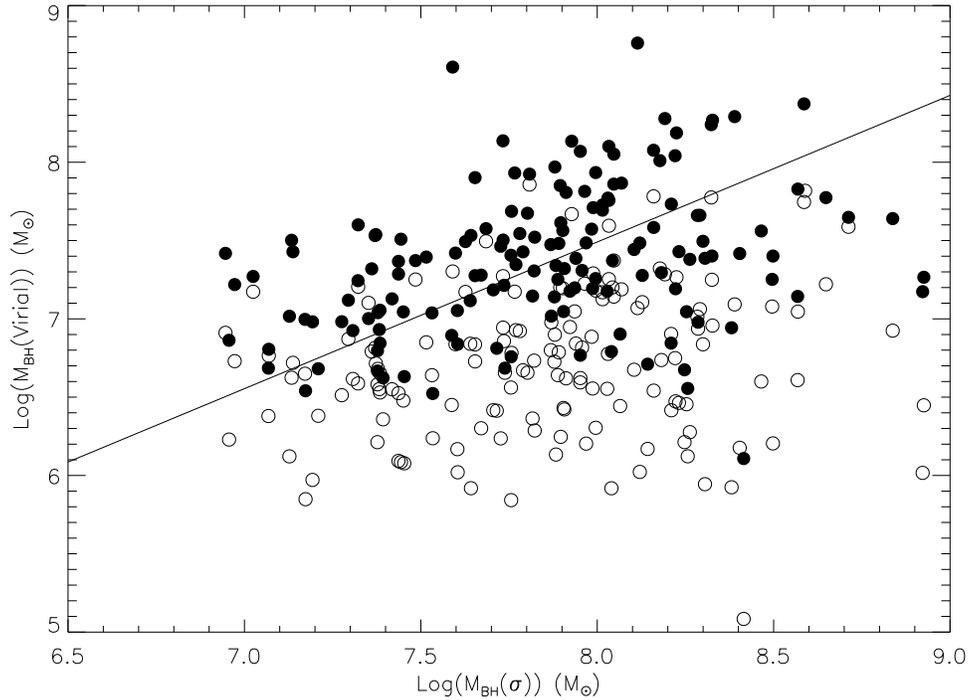}
\caption{The correlation between the two kinds of BH masses, $M_{BH}(\sigma)$
estimated from Equation (1) and $M_{BH}(Virial)$ estimated from Equation (2).
The solid line represents the relation: $\log(M_{BH}(\sigma))=1.06\log(M_{BH}(Virial))$.
Solid circles represent calculations after the internal reddening correction, open circles are the values
before the internal reddening correction.}
\end{figure}

\begin{figure}
\centering\includegraphics[width=140mm,height = 100mm]{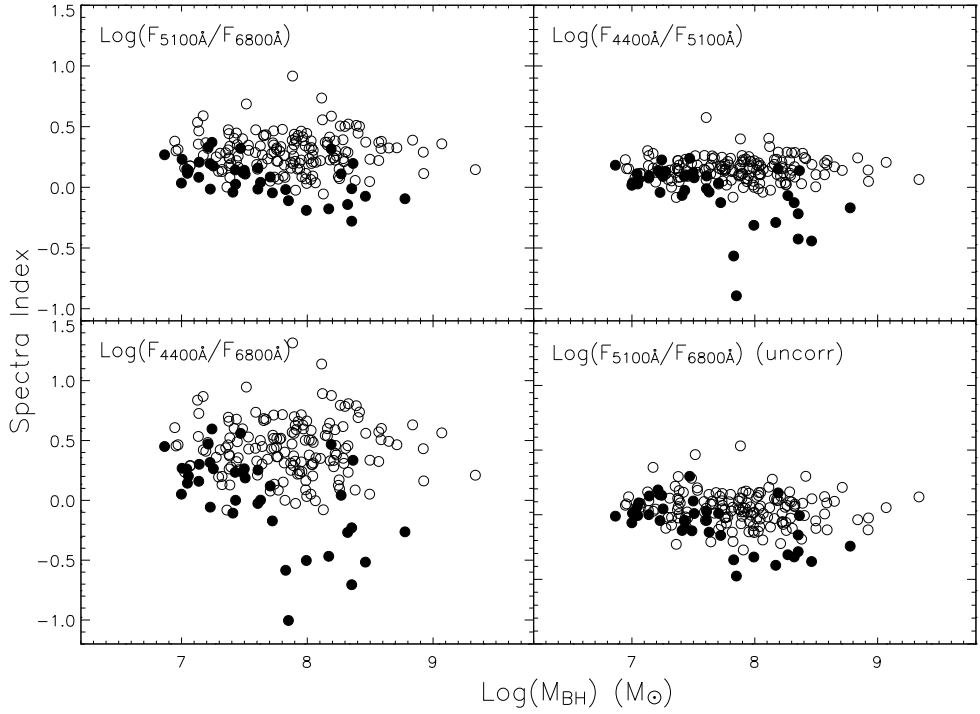}
\caption{The correlation between the spectral index and the central
BH mass. Solid circles represent the low luminosity AGN with luminosity of H$\alpha$ less than $10^{41} {\rm erg\cdot s^{-1}}$ as identified in Ho et al. (1997a, 1997b). Open circles are the normal AGN.}
\end{figure}

\begin{figure}
\centering\includegraphics[width=140mm,height = 100mm]{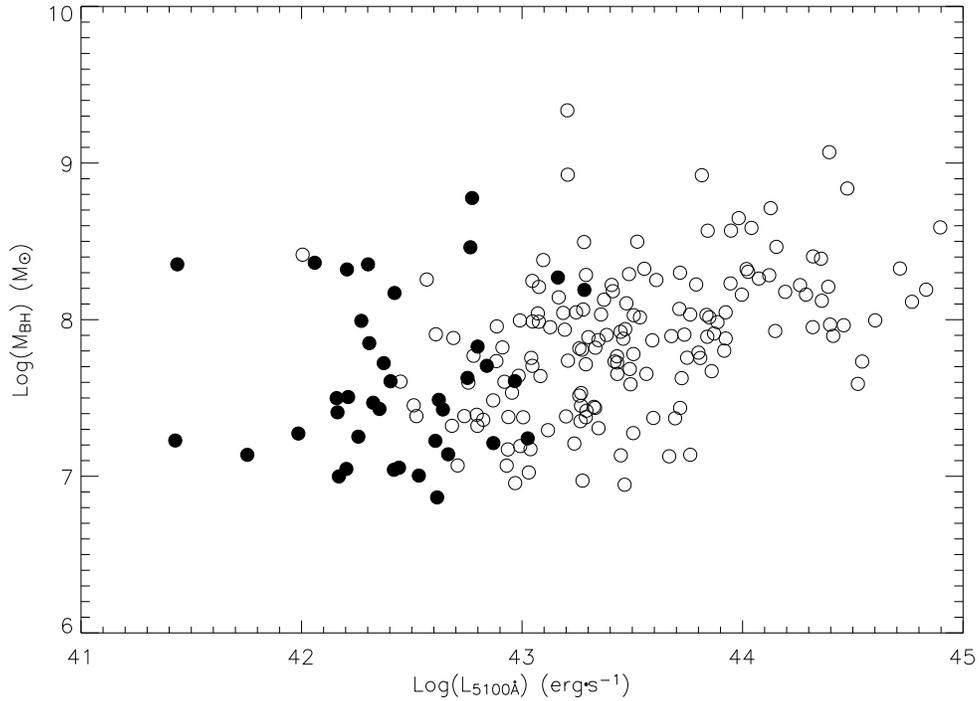}
\caption{The correlation between the central BH mass and the continuum
luminosity after the correction of internal reddening effects. Solid circles represent the low luminosity AGN with luminosity of H$\alpha$ less than $10^{41} {\rm erg\cdot s^{-1}}$ as identified in
Ho et al. (1997a, 1997b),open circles
are the normal AGN.}
\end{figure}

\begin{figure}
\centering\includegraphics[width=140mm,height = 100mm]{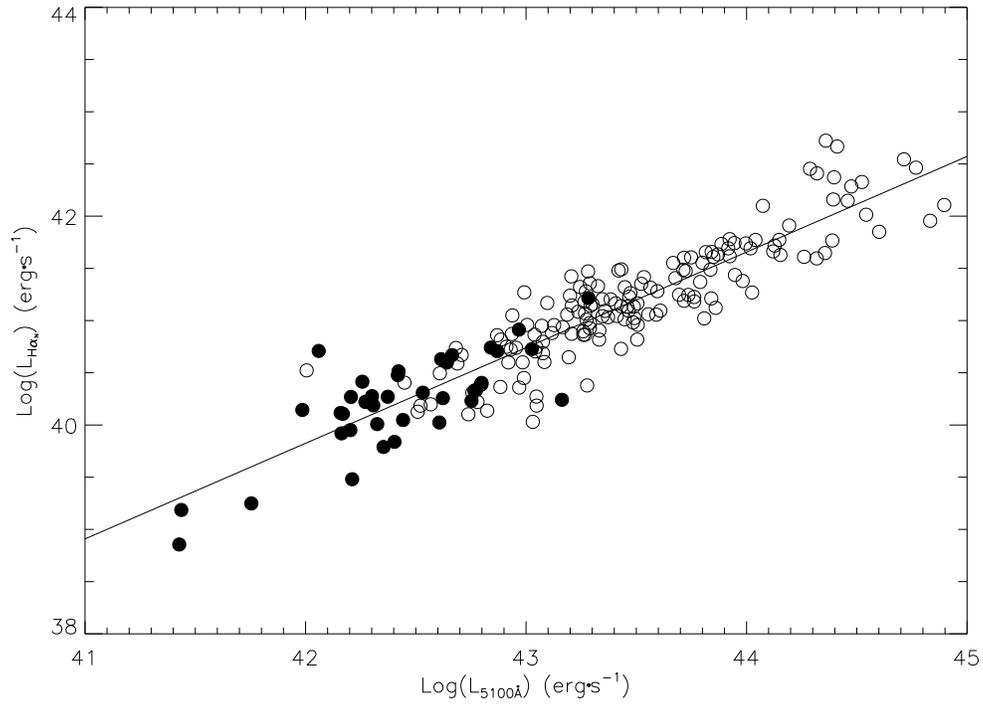}
\caption{The correlation between the continuum luminosity and the luminosity
of narrow H$\alpha$ after the correction of internal reddening effects. Solid circles represent the low luminosity AGN with luminosity of H$\alpha$ less than $10^{41} {\rm erg\cdot s^{-1}}$ as identified in Ho et al. (1997a, 1997b), open circles are the normal AGN. The solid line represents the best fit $L_{H\alpha_N}\propto L_{5100\AA}^{0.915\pm0.003}$.}
\end{figure}

\begin{thebibliography}{}
\bibitem[Adelman-McCarthy et al., 2006]{am06}
Adelman-McCarthy J. K., et al., 2006, ApJS, 162, 38
\bibitem[Antonucci, 1993]{an93}
Antonucci R., 1993, ARA\&A, 31, 473
\bibitem[Bonning et al., 2006]{bo06}
Bonning E. W., Shields G. A., Salviander S., Cheng L., Gebhardt K., 2007,
ApJ, 659, 211
\bibitem[Dultzin-Hacyan et al., 2007]{du07}
Dultzin-Hacyan D., Marziani P., Sulentic J. W., 2007, Black Holes: From Stars to Galaxies - Across the Range of Masses, International Astronomical Union. Symposium no. 238, held 21-25 August, 2006 in Prague, Czech Republic, 238, 13
\bibitem[Ferrarese \& Merritt, 2000]{fm00}
Ferrarese L., Merritt D., 2000, ApJ, 539, L9
\bibitem[Fitzpatrick, 1999]{fi99}
Fitzpatrick E. L., 1999, PASP, 111, 63
\bibitem[Francis et al. 1991]{fr91}
Francis P. J., Hewett P. C., Foltz C. B., et al., 1991, ApJ, 373, 465
\bibitem[Gebhardt et al., 2000]{gb00}
Gebhardt K., et al., 2000, ApJ, 539, L13
\bibitem[Green \& Ho, 2005a]{gh05a}
Greene J. E., Ho L. C., 2005a, ApJ, 627, 721
\bibitem[Green \& Ho, 2005b]{gh05b}
Greene J. E., Ho L. C., 2005b, ApJ, 630, 122
\bibitem[Hao et al., 2005]{hao05}
Hao L., et al., 2005, AJ, 129, 1783
\bibitem[H\"{a}ring \& Rix, 2004]{hr04}
H\"{a}ring, N., Rix, Hans-Walter, 2004, ApJ, 640, L89
\bibitem[Ho et al., 1997a]{ho97a}
Ho L. C., Filippenko A. \& Sargent W. L. W., 1997a, ApJS, 112, 315
\bibitem[Ho et al., 1997b]{ho97b}
Ho L. C., Filippenko A. \& Sargent W. L. W., 1997b, ApJS, 112, 391
\bibitem[Ho, 1999]{ho99}
Ho L. C., 1999, ApJ, 516, 672
\bibitem[Kaspi et al., 2000]{kas00}
Kaspi S., Smith P. S., Netzer H., Maoz D., Jannuzi B. T., Giveon U., 2000,
ApJ, 533, 631
\bibitem[Kaspi et al., 2005]{kas05}
Kaspi S., Maoz D., Netzer H., Peterson B. M., Vestergaard M., Jannuzi B. T., 2005, ApJ, 629, 61
\bibitem[Kauffmann et al., 2003]{kg03}
Kauffmann G., et al., 2003, MNRAS, 346, 1055
\bibitem[Kormendy, 2001]{ko01}
Kormendy J., 2001, in Galaxy Disks and Disk Galaxies, proceeding of a conference held in Rome, Italy, June 12-16, 2000 at the Pontifical Gregorian University and sponsored by the Vatican Observatory. ASP Conference Series, Vol. 230. Edited by Jose G. Funes, S. J. and Enrico Maria Corsini. San Francisco: Astronomical Society of the Pacific. ISBN: 1-58381-063-3, 2001, pp. 247-256
\bibitem[Laor, 2001]{l01}
Laor A., 2001, ApJ, 533, 677
\bibitem[Borgne et al., 2003]{lb03}
Le Borgne J. F., et al., 2003, A\&A, 402, 433
\bibitem[Li et al., 2005]{li05}
Li C., Wang T. G., Zhou H. Y., Dong X. B., Chen F. Z., 2005, AJ, 129, 669
\bibitem[Marconi \& Hunt, 2003]{mh03}
Marconi A., Hunt L. K., 2003, ApJ, 589, L21
\bibitem[McLure \& Dunlop, 2002]{md02}
McLure R. J., Dunlop J. S., 2002, MNRAS, 331, 795
\bibitem[Merritt \& Ferrarese, 2001]{fm01}
Merritt D., Ferrarese L., 2001, ApJ, 547, 140
\bibitem[Netzer, 2003]{n03}
Netzer H., 2003, ApJ, 583, L5
\bibitem[Onken et al., 2004]{on04}
Onken C. A., Ferrarese L., Merritt D., Peterson B. M., Pogge R. W., Vestergaard M., Wandel A., 2004, ApJ, 615, 645
\bibitem[Panessa et al., 2006]{pa06}
Panessa F., Bassani L., Cappi M., Dadina M., Barcons X., Carrera F. J., Ho L. C., Iwasawa K., 2006, A\&A, 455, 173
\bibitem[Peterson et al., 2004]{pf04}
Peterson B. M., et al., 2004, ApJ, 613, 682
\bibitem[Quintilio \& Viegas, 1997]{qv97}
Quintilio, R.; Viegas S. M., 1997, ApJ, 474, 616
\bibitem[Rix \& White, 1992]{rw92}
Rix H.-W., White S. D. M., 1992, MNRAS, 254, 389
\bibitem[Shang et al., 2005]{sh05}
Shang Z., Brotherton M. S., Green R. F., Kriss G. A., Scott J., Quijano J. K.,
2005, ApJ, 619, 41
\bibitem[Sulentic et al., 2006]{su06}
Sulentic J. W., Repetto P., Stirpe G. M., Marziani P., Dultzin-Hacyan D.,
Calvani M., 2006, A\&A, 456, 929
\bibitem[Tremaine et al., 2002]{tr02}
Tremaine S., Gebhardt K., Bender R., Bower G., et al., 2002, ApJ, 574,
740
\bibitem[Urry \& Padovani, 1995]{up95}
Urry C. M., Padovani P., 1995, PASP, 107, 803
\bibitem[Berk et al. 2001]{berk01}
Vanden Berk, D. E., et al., 2001, AJ, 122, 549
\bibitem[Wandel, 1999]{w99}
Wandel A., 1999, ApJ, 519, L39
\bibitem[Wang \& Zhang, 2003]{wa03}
Wang T.-G., Zhang X.-G., 2003, MNRAS, 340, 793
\bibitem[Zhang et al., 2007a]{zh07a}
Zhang X.-G., Dultzin-Hacyan D., Wang T.-G., 2007a, MNRAS, 374, 691
\bibitem[Zhang et al., 2007b]{zh07b}
Zhang X.-G., Dultzin-Hacyan D., Wang T.-G., 2007b, MNRAS, 376, 1335
\bibitem[Zhang et al., 2007c]{zh07c}
Zhang X.-G., Dultzin-Hacyan D., Wang T.-G., 2007c, MNRAS, 377, 1215
\bibitem[Zheng et al. 1997]{zh97}
Zheng W., Kriss G. A., Telfer R. C., Grimes J. P., Davidsen A. F., 1997, ApJ,
475, 469
\end{thebibliography}
\end{document}